# Multitasking additional-to-driving: Prevalence, structure, and associated risk in SHRP2 naturalistic driving data

András Bálint[1,4], Carol A. C. Flannagan[2,1], Andrew Leslie[2], Sheila Klauer[3], Feng Guo[3], Marco Dozza[1,4]

[1]Chalmers University of Technology, Vehicle Safety Division, Gothenburg, Sweden
[2]University of Michigan Transportation Research Institute, Ann Arbor, MI, USA
[3]Virginia Polytechnic and State University, Blacksburg, VA, USA
[4]SAFER Vehicle and Traffic Safety Centre at Chalmers, Gothenburg, Sweden

**ABSTRACT**

Objective: This paper 1) analyzes the extent to which drivers engage in multitasking additional-to-driving (MAD) under various conditions, 2) specifies odds ratios (ORs) of crashing associated with MAD, and 3) explores the structure of MAD.

Methods: Data from the Second Strategic Highway Research Program Naturalistic Driving Study (SHRP2 NDS) was analyzed to quantify the prevalence of MAD in normal driving as well as in safety-critical events of various severity level and compute point estimates and confidence intervals for the corresponding odds ratios estimating the risk associated with MAD compared to no task engagement. Sensitivity analysis in which secondary tasks were re-defined by grouping similar tasks was performed to investigate the extent to which ORs are affected by the specific task definitions in SHRP2. A novel visual representation of multitasking was developed to show which secondary tasks co-occur frequently and which ones do not.

Results: MAD occurs in 11% of control driving segments, 22% of crashes and near-crashes (CNC), 26% of Level 1-3 crashes and 39% of rear-end striking crashes, and 9%, 16%, 17% and 28% respectively for the same event types if MAD is defined in terms of general task groups. The most common co-occurrences of secondary tasks vary substantially among event types; for example, "Passenger in adjacent seat – interaction" and "Other non-specific internal eye glance" tend to co-occur in CNC but tend not to co-occur in control driving segments. The odds ratios of MAD using SHRP2 task definitions compared to driving without any secondary task and the corresponding 95% confidence intervals are 2.38 (2.17-2.61) for CNC, 3.72 (3.11-4.45) for Level 1-3 crashes and 8.48 (5.11-14.07) for rear-end striking crashes. The corresponding ORs using general task groups to define MAD are slightly lower at 2.00 (1.80-2.21) for CNC, 3.03 (2.48-3.69) for Level 1-3 crashes and 6.94 (4.04-11.94) for rear-end striking crashes.

Conclusions: The number of secondary tasks that the drivers were engaged in differs substantially for different event types. A graphical representation was presented that allows mapping task prevalence and co-occurrence within an event type as well as a comparison between different event types. The ORs of MAD indicate an elevated risk for all safety-critical events, with the greatest increase in the risk of rear-end striking crashes. The results are similar independently of whether secondary tasks are defined according to SHRP2 or general task groups. The results confirm that the reduction of driving performance from MAD observed in simulator studies is manifested in real-world crashes as well.

**ACKNOWLEDGMENTS**

We would like to thank the Transportation Research Board and the National Academies for making the SHRP2 data available for analysis. The findings and conclusions of this paper are those of the author(s) and do not necessarily represent the views of VTTI, the Transportation Research Board, or the National Academies.

This research was sponsored by the Alliance of Automotive Manufacturers. The opinions expressed within this paper represent those of the authors and not necessarily those of the funding agency.

# 1 INTRODUCTION

Driver distraction has a widespread research literature indicating that it is a major contributing factor in road crashes around the world. Reviews of the topic include Young & Regan (2007), Regan, Lee, & Young (2009), and Papantoniou, Papadimitriou, & Yannis (2017). Various research methods and data sources have been used to study the consequences of driver distraction, including simulator studies and other experimental designs. The analysis of data from large-scale naturalistic driving studies has provided valuable results regarding the real-world risk associated with a range of secondary tasks. Starting from the analysis of the 100-Car data in Klauer et al. (2006), different naturalistic driving databases have been studied to estimate crash risk and near-crash risk, including, but not limited to, recent studies of the Second Strategic Highway Research Program Naturalistic Driving Study data from the United States (Dingus, et al., 2016; Guo, et al., 2017; Flannagan, Bärgman, & Bálint, 2019) and European data from UDRIVE (Ismaeel, Hibberd, & Carsten, 2018; Morgenstern, Naujoks, Krems, & Keinath, 2018).

Nonetheless, a substantial knowledge gap in the literature of driver distraction identified in Lansdown, Stephens, & Walker (2015) is that very few studies specifically address multiple driver distractions or *multitasking additional-to-driving (MAD)*. In this reference, a systematic review of relevant studies, which are mainly simulator studies, is made and it is found that MAD is "almost universally detrimental to driving performance".

The aim of this paper is to specifically analyze multiple driver distractions based on naturalistic driving data. For this purpose, the following objectives are addressed: 1) quantifying the extent to which drivers are engaged in MAD under various conditions; 2) computing point estimates and confidence intervals for crash odds ratios of MAD; 3) exploring the structure of multitasking; and 4) investigating the sensitivity of results with respect to the definition of MAD in terms of more general tasks.

This paper is structured as follows. The analysis methods addressing the objectives specified above are described in Section 2, followed by the corresponding results in Section 3. The results include several graphs displaying the structure of MAD; some of these are provided on separate pages in an Appendix, for better readability. A discussion of the results and description of limitations is provided in Section 4 and the conclusions of this study are summarized in Section 5.

# 2 METHODS

This section includes a description of the data used for the analysis and the definition of MAD as used in this paper (Section 2.1), together with the description of graphs used for visualization of the structure of MAD (Section 2.2) and how risk associated with MAD is quantified in terms of various odds ratios (Section 2.3).

## 2.1 The SHRP2 naturalistic driving database

Data from the Second Strategic Highway Research Program Naturalistic Driving Study (SHRP2 NDS) is used for this study. The full SHRP2 sample, collected from October 2010 to December 2013, includes 3,546 drivers from six locations around the U.S., ranging in age from 16 to 98 years, with young and old drivers (aged 16-24, respectively 65+) overrepresented compared to the general U.S. driver population. Further details regarding data collection and data coding in SHRP2 may be found in Kidd & McCartt (2015) and Hankey, Perez, & McClafferty (2016).

### 2.1.1 Event types

The SHRP2 dataset includes over 37,000 events in total, classified as crashes, near-crashes and baseline events. Crashes are divided into the following four severity levels: *Level 1 Severe Crash; Level 2 Crash Moderate Severity; Level 3 Crash Minor Severity;* and *Level 4 Crash Tire strike, low risk.* A near-crash is any circumstance requiring a rapid evasive maneuver by the subject vehicle or any other vehicle, pedestrian, cyclist, or animal to avoid a crash. Baseline events include both "balanced-sample" baselines, which were selected at random from each participant in proportion to time driven, and "additional" baselines that do not satisfy the balancing requirement. For more detailed definitions and other relevant aspects of the SHRP2 dataset (e.g. intra-rater reliability scores), see Hankey, Perez, & McClafferty (2016). This analysis excludes events with unknown secondary task engagement and focuses primarily on the following event types:

- *Control driving segments (CDS)*: Balanced-sample baseline events, i.e. baseline events stratified by participant, proportional to hours traveled and limited to driving over 5 mph;
- *Crashes and near-crashes (CNC)*: Crashes of all levels plus near-crashes;
- *Level 1-3 crashes (L1-3)*: All crashes excluding low-risk tire strikes;
- *Rear-end striking crashes (Rear-end striking)*: Rear-end crashes of severity level 1-3 in which the vehicle participating in SHRP2 NDS was the striking vehicle.

These event types have been considered in several other analyses of naturalistic driving data, e.g. in Klauer et al. (2006), Victor et al. (2014), Kidd & McCartt (2015), Dingus et al. (2016), Guo et al. (2017). Additionally, following the approach in Flannagan et al. (2018), the following five event types, which all can be viewed as variations of level 1-3 crashes, are also considered for a subset of analyses:

- *Level 1-2 crashes (L1-2)*: This includes crashes of severity levels 1 and 2; according to Hankey, Perez, & McClafferty (2016), these are crashes that are likely to be police-reported;
- *All crashes (L1-4)*: Crashes of all severities including low-risk tire strikes;
- *At-fault level 1-3 crashes (L1-3 at-fault)*: Crashes of severity level 1-3 in which the driver was assigned to be at-fault;
- *Level 1-3 crashes plus near-crashes (L1-3+NC)*: CNC excluding low-risk tire strikes;
- *Run-off-road crashes (Run-off-road)*: Run-off-road events of crash severity level 1-3.

Overall, nine event types have been defined above, including CDS and eight other types that will be referred to as *safety-critical event types*. The following section provides details regarding the coding of secondary task engagement within events of different types.

### 2.1.2  Data coding and definitions

The coding of secondary tasks is performed by data reductionists with specialized training, based on continuous video recordings. For control driving segments, a 6-second epoch is selected at random and involvement in up to three secondary tasks within the 6-second time frame is coded according to the definitions in the data dictionary (Hankey, Perez, & McClafferty, 2016). For crashes or near-crashes, the coding was designed to evaluate the sequence of events that occurred in the seconds prior to crashes and near-crashes. In each crash or near-crash event, secondary task engagement is coded 5 seconds prior a precipitating event until the conflict ends. For ease of reference, the time frame considered for coding secondary tasks will be called the *event window* in this paper.

*Multitasking additional-to-driving (MAD)* is defined to be present for an event if engagement in at least two secondary tasks is coded for the event. Secondary task engagement is coded if it occurred any time during the event window, so engagement in the secondary tasks for multitasking can be consecutive or concurrent.

*Single task additional-to driving (SAD)* is defined to be present if there is exactly one secondary task coded for an event. Consequently, SAD and MAD comprise a mutually exclusive and collectively exhaustive partitioning of *secondary task engagement* (i.e., one or more tasks).

*Engagement in at least 3 tasks* (abbreviated as *3+ tasks*) is a specific part of MAD which is also considered in this study.

The *prevalence* of events satisfying a given property within a specific event type (e.g. events with MAD within control driving segments) is the proportion of events satisfying the given property compared to all events within the event type. Additionally, for the graphs described in Section 2.2, prevalence of secondary task co-occurrence for tasks A and B will be computed as the number of events with A and B co-occurring in comparison to the event counts with A or B, respectively. The resultant pair of values can differ, such as when 50% of events with A include co-occurrence with B, but only 10% of events with B include co-occurrence with A.

### 2.1.3  General task definitions for sensitivity analysis

In Section 2.1.2, MAD is defined in terms of secondary tasks definitions used in the SHRP2 database, and the corresponding results on MAD are therefore specific to this coding. For example, SHRP2 uses a detailed coding scheme in defining secondary tasks (e.g. different subtasks related to cell-phone use are coded separately) leading to a higher rate of defining events as involving MAD relative to other studies. Additional analysis was performed to investigate how the results change if more general secondary tasks are considered. To explore the effect of secondary task definitions on the prevalence of MAD and associated risk, a sensitivity analysis was performed using general secondary tasks defined as combinations of secondary tasks in SHRP2, see Table 1 below. Note that the 14 general secondary tasks in Table 1, denoted as G14 in this paper, are deliberately defined to be as general as possible for investigating an extreme case in the sensitivity analysis in that any instance of MAD in terms of G14 tasks corresponds to engagement in at least two secondary tasks that are from substantially different task groups. The set G14 is defined for this specific purpose and is not meant to represent an appropriate set of secondary tasks for other analyses.

**Table 1 Definition of G14, i.e. 14 general secondary tasks for sensitivity analysis.**

| # | Secondary tasks in SHRP2 included in the general task |
|---|---|
| 1 | Talking/singing, audience unknown; Dancing |
| 2 | Reading; Writing |
| 3 | Passenger in adjacent seat – interaction; Passenger in rear seat – interaction; Child in adjacent seat – interaction; Child in rear seat – interaction |
| 4 | Moving object in vehicle; Insect in vehicle; Pet in vehicle; Object dropped by driver; Reaching for object, other; Object in vehicle, other |
| 5 | Cell phone, Holding; Cell phone, Talking/listening, hand-held; Cell phone, Talking/listening, hands-free; Cell phone, Texting; Cell phone, Browsing; Cell phone, Dialing hand-held; Cell phone, Dialing hand-held using quick keys; Cell phone, Dialing hands-free using voice-activated software; Cell phone, Locating/reaching/answering; Cell phone, other |
| 6 | Tablet device, Locating/reaching; Tablet device, Operating; Tablet device, Viewing; Tablet device, Other |
| 7 | Adjusting/monitoring climate control; Adjusting/monitoring radio; Inserting/retrieving CD (or similar); Adjusting/monitoring other devices integral to vehicle |
| 8 | Looking at previous crash or incident; Looking at pedestrian; Looking at animal; Looking at an object external to the vehicle; Distracted by construction; Other external distraction |
| 9 | Reaching for food-related or drink-related item; Eating with utensils; Eating without utensils; Drinking with lid and straw; Drinking with lid, no straw; Drinking with straw, no lid; Drinking from open container |
| 10 | Reaching for cigar/cigarette; Lighting cigar/cigarette; Smoking cigar/cigarette; Extinguishing cigar/cigarette |
| 11 | Reaching for personal body-related item; Combing/brushing/fixing hair; Applying make-up; Shaving; Brushing/flossing teeth; Biting nails/cuticles; Removing/adjusting clothing; Removing/adjusting jewelry; Removing/inserting/ adjusting contact lenses or glasses; Other personal hygiene |
| 12 | Other non-specific internal eye glance |
| 13 | Other known secondary task |
| 14 | Unknown type (secondary task present) |

The general tasks defined in Table 1 are used for comparisons of prevalence and risk of MAD under different task definitions and are not used for the visualization of secondary task co-occurrences as detailed in Section 2.2, which are exclusively based on the SHRP2 secondary task definitions.

## 2.2 Visualization of the structure of MAD

Beyond the quantification of the prevalence of MAD, our aim is to understand its structure, i.e. investigate which secondary tasks tend to co-occur and which secondary tasks do not tend to co-occur. A visualization of the results using graph plots was developed in order to give an immediate idea about these aspects and aid comparisons between event types.

Secondary task prevalence within an event type was visualized by points called *nodes,* where point size is proportional to prevalence. Lines called *edges* were placed between nodes and line width is proportional to the frequency of co-occurrence of the corresponding two secondary tasks. Due to the large number of secondary tasks and possible co-occurrences, various constraints were introduced to simplify the graphs to highlight the most prevalent secondary tasks and particularly common co-occurrences (relative to the prevalence of the corresponding secondary tasks). Nodes and edges in the graphs are labeled to express the prevalence of secondary tasks and co-occurrences; the details of labeling will be described in Section 3.2.

Additionally, different line styles and a color coding of edges were introduced to represent how often two secondary tasks co-occur compared to the expected frequency under an independence model. Dark green, solid edges represent significantly higher co-occurrence than expected whereas dark red, dashed edges represent significantly lower co-occurrence. In both cases, significance is determined using chi-squared test of independence ($p < 0.05$). Blue, dash-dotted edges indicate that the difference between observed and expected frequency of co-occurrence is not statistically significant. Edges for which the chi-squared test is not performed due to an expected frequency count below 5 but have a correlation coefficient $r > 0.1$ are indicated by a light green solid line while those with an expected frequency < 5 and $r < -0.1$ are indicated by a light red dashed line.

With small modifications, the developed method can be used for the comparison of secondary task prevalence and co-occurrence between different event types. For making a visual comparison, it is important that the graphs constructed for the different event types contain the same set of secondary tasks (which is not a given based on the rules above, as not all secondary tasks may be displayed due to the simplification of the graph), and that the nodes representing the tasks are in the same geometrical position.

## 2.3 Risk associated with MAD

This section describes the computation of various estimates that quantify the safety risk associated with MAD. All measures are *odds ratios (ORs)* whose computation is based on a cross-tabulation of the number of *cases* and *controls* for events including driver behaviors whose associated risk is assessed and a reference level, see Table 2 below. As shown in Guo (2019), the odds ratio based on randomly sampled control driving segments is an approximation of the crash rate ratio under the distracted condition and the reference condition.

Table 2 Cross-tabulation of event counts used for odds ratio calculation.

|  | Cases | Controls |
|---|---|---|
| Driver behavior present | a | b |
| Reference level | c | d |

The cell counts depend on the precise definitions of the row and column headers in Table 2. One particular choice considered in this study is the following: "Cases" are L1-3 (i.e., level 1-3 crashes); "Controls" are CDS (this is the same for all ORs considered in this paper); "Driver behavior" is MAD and "Reference level" is driving without secondary task engagement. Given specific definitions, the *crude odds ratio* can be computed by dividing the ratio of cases to controls when the driver behavior is present by the corresponding ratio for the reference level:

$$\widehat{OR} = \frac{a/b}{c/d}.$$

The crude odds ratio approximates the risk associated with the driver behavior because safety-critical events are rare events. The endpoints of the corresponding 95% confidence intervals are computed as follows:

$$exp\left(\log(\widehat{OR}) \pm 1.96 * \sqrt{\frac{1}{a} + \frac{1}{b} + \frac{1}{c} + \frac{1}{d}}\right).$$

Computing $\widehat{OR}$ with the definitions as in the example above estimates the risk of level 1-3 crashes when engaged in MAD in comparison to driving without secondary task engagement, and the confidence interval quantifies the uncertainty about the estimate.

Other event types as defined in Section 2.1.1 will also be addressed besides level 1-3 crashes; additionally, apart from MAD as driver behavior, SAD and 3+ tasks will also be considered. These choices are expressed in the notation as subscripts (case definition) and superscripts (behavior definition and reference level); e.g., the OR computed with the definitions as in the example above is denoted as $\widehat{OR}_{L1-3}^{MAD/No\ Tasks}$. Note that the ratio $\widehat{OR}_{L1-3}^{MAD/No\ Tasks} / \widehat{OR}_{L1-3}^{SAD/No\ Tasks}$ is also an odds ratio, with level 1-3 crashes as case definition, MAD as driver behavior and SAD as reference level and is an estimate of the risk associated with adding further tasks to a single secondary task additional-to-driving. Analogous statements hold for $\widehat{OR}_{L1-3}^{3+\ tasks/No\ Tasks} / \widehat{OR}_{L1-3}^{SAD/No\ Tasks}$ and $\widehat{OR}_{L1-3}^{3+\ tasks/No\ Tasks} / \widehat{OR}_{L1-3}^{MAD/No\ Tasks}$.

## 3 RESULTS

### 3.1 Prevalence of multitasking additional-to-driving

The first result, presented in Figure 1 below, is the quantification of the prevalence of MAD and SAD using SHRP2 task definitions across all event types defined in Section 2.1.1, with the event types ordered by the prevalence of MAD. The y-axis represents the percentage of events of the given type.

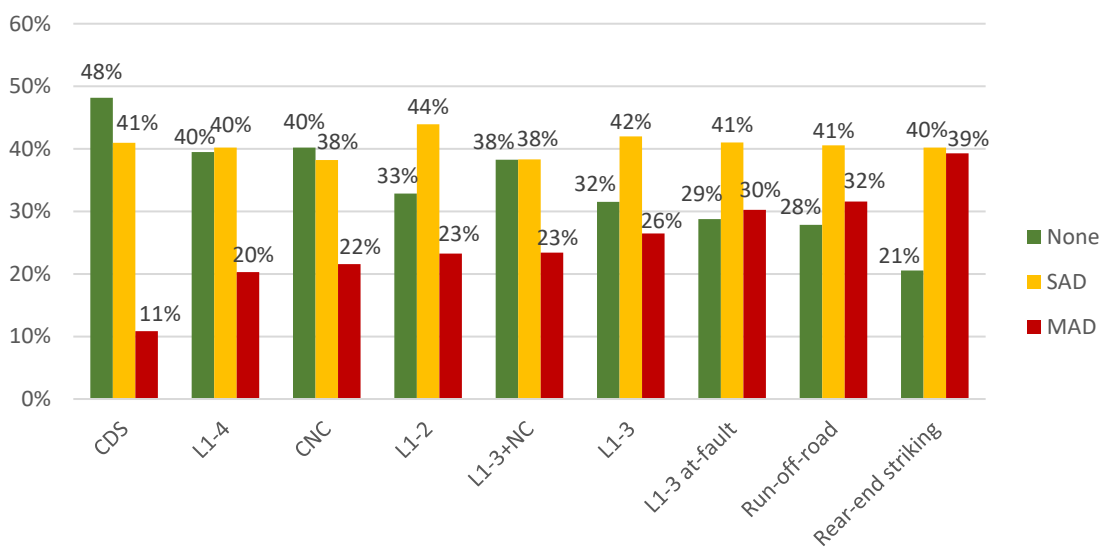

Figure 1 Prevalence of secondary task engagement across event types using SHRP2 task definitions.

It is immediately clear from this figure that the number of secondary tasks that the drivers were engaged in differs substantially for different event types. For example, the prevalence of secondary task engagement (i.e. SAD plus MAD) is 52% for control driving segments, 60% for crashes and near-crashes, 68% for Level 1-3 crashes and 79% for rear-end striking crashes, while the prevalence of MAD is 11% for CDS, 22% for CNC, 26% for Level 1-3 crashes and 39% for rear-end striking crashes. At the same time, the prevalence of SAD is almost constant at around 40% across all event types.

The corresponding graph based on general task definitions (G14, see Table 1) is presented in Figure 2 below. Using G14 does not change the prevalence of secondary task engagement, but the proportion of events with MAD compared to SAD is somewhat lower than in Figure 1. Note that the order of events by the prevalence of MAD is unchanged by the usage of general task definitions. For all graphs considered in this paper, bars corresponding to G14 tasks will be marked by a pattern of diagonal stripes for easier identification.

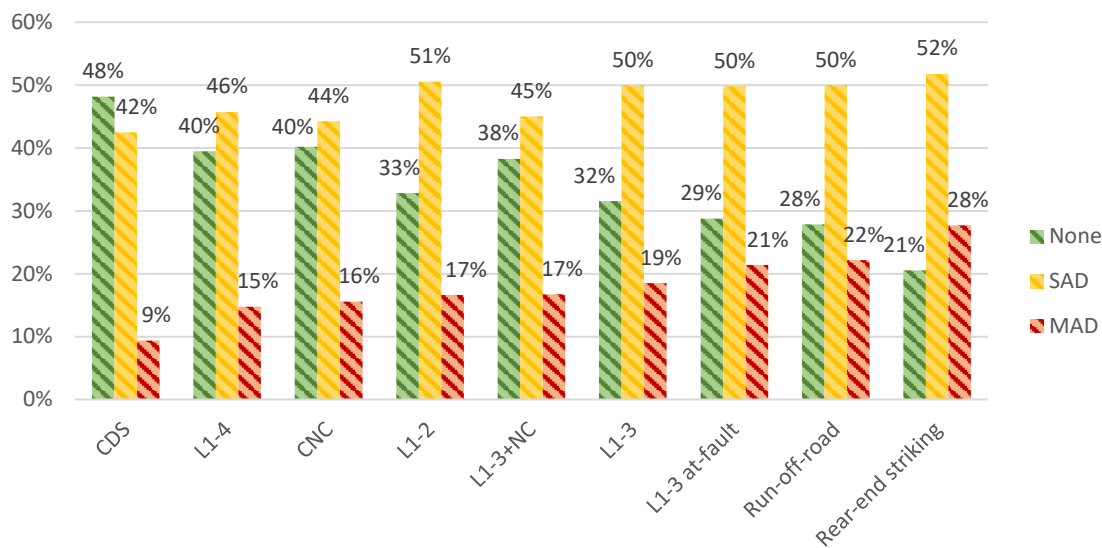

Figure 2 Prevalence of secondary task engagement across event types using G14 tasks.

Finally, Table 3 summarizes the prevalence of MAD along with the prevalence of being engaged in at least three secondary tasks for the different secondary task definitions. With both secondary task definitions considered, both MAD and engagement in at least three tasks have the smallest relative frequency among control driving segments and the highest relative frequency among rear-end striking crashes. The differences between event types are somewhat smaller using the general task definitions: multitasking is 1.6-3 times more common and engagement in at least three tasks is 1.8-3.6 times more common among other event types than among control driving segments, while the corresponding multiplier ranges are 1.9-3.6 and 2.9-7 using SHRP2 definitions.

Table 3 Prevalence of MAD and engagement in at least three tasks for both task definitions.

|                                      |                   | SHRP task definitions | | G14 tasks | |
| ------------------------------------ | ----------------- | --- | -------- | --- | -------- |
| Event type                           | Number of events  | MAD | 3+ tasks | MAD | 3+ tasks |
| Control driving segments             | 19991             | 11% | 2%       | 9%  | 1%       |
| All crashes                          | 1524              | 20% | 4%       | 15% | 2%       |
| Crashes and near-crashes             | 4224              | 22% | 6%       | 16% | 2%       |
| Level 1-2 crashes                    | 271               | 23% | 6%       | 17% | 3%       |
| Level 1-3 crashes plus near-crashes  | 3613              | 23% | 6%       | 17% | 2%       |
| Level 1-3 crashes                    | 891               | 26% | 6%       | 19% | 2%       |
| At-fault level 1-3 crashes           | 622               | 30% | 7%       | 21% | 2%       |
| Run-off-road                         | 402               | 32% | 6%       | 22% | 2%       |
| Rear-end striking                    | 112               | 39% | 11%      | 28% | 4%       |

## 3.2 Visualization of secondary task co-occurrence

A graph representing prevalence and co-occurrence of all secondary tasks for control driving segments is shown in Figure 3. This graph includes all secondary tasks and co-occurrences, but only shows the labels of nodes with case count >50 within CDS for readability. This graph illustrates the complexity of relationships and justifies the need to use cutoff thresholds for further analysis, furthermore shows that there are relatively few solid (green) edges indicating significantly more frequent co-occurrence than assuming independent occurrence compared to dash-dotted (blue), dashed (red) or missing edges. The labels on nodes indicate the name of the secondary task represented by the node and the percentage and number of events with the given secondary task. For example, the secondary task corresponding to the second largest node, "Other external distraction," is present in 2021 control driving segments, which is approximately 10% of the total number of 19991 control driving segments.

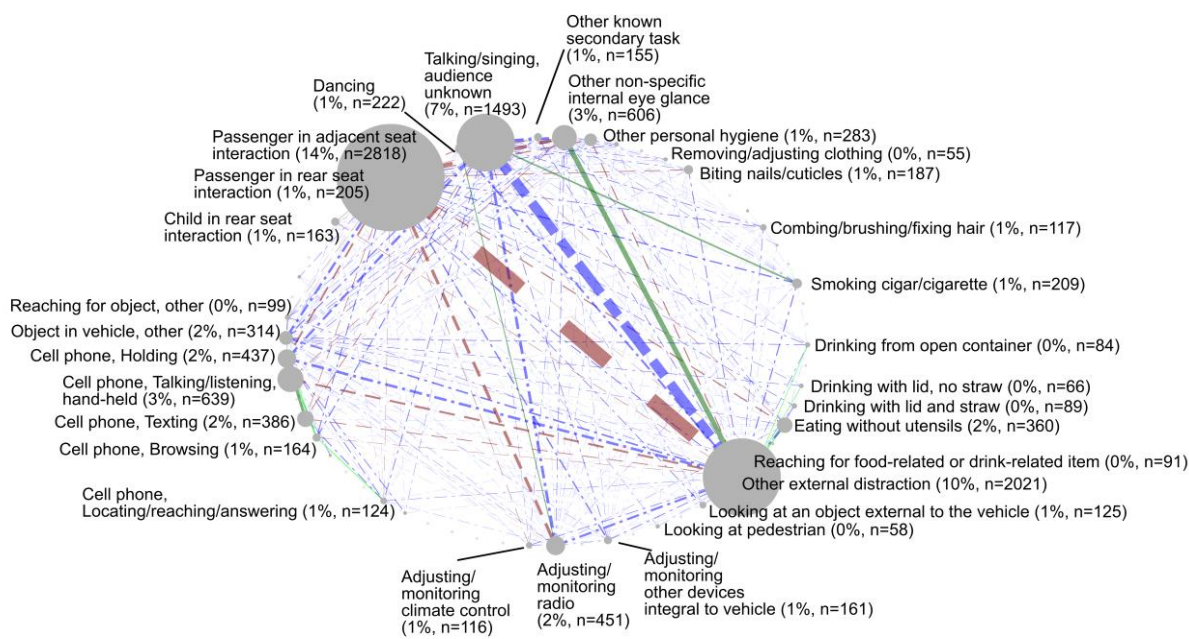

Figure 3 Co-occurrence graph for control driving segments (n=19991). Edge labels and the labels of nodes with case count below 50 are not shown. Green (solid) edges indicate greater frequency, red (dashed) edges indicate smaller frequency than independent co-occurrence, and blue (dash-dotted) edges indicate no significant difference compared to independent co-occurrence.

Figure 4 is a simplified version of Figure 3 and the labels on edges indicate the number of events of the given type in which the secondary tasks corresponding to the end-nodes of the edge co-occur, as well as the relative frequency of this co-occurrence compared to the prevalence of the tasks. For example, the tasks "Cell phone, Talking/listening, hand-held," (the bottom-left node in Figure 4) and "Other external distraction" (the lower node at the right side of Figure 4) having counts of 639 and 2021 respectively (i.e., CDS prevalence of 3% and 10%), co-occur in 32 control driving segments. As 32/639 = 5% and 32/2021 = 2% and relative frequencies are displayed in ascending order, the label on the edge between these secondary tasks is "32 (2%; 5%)."

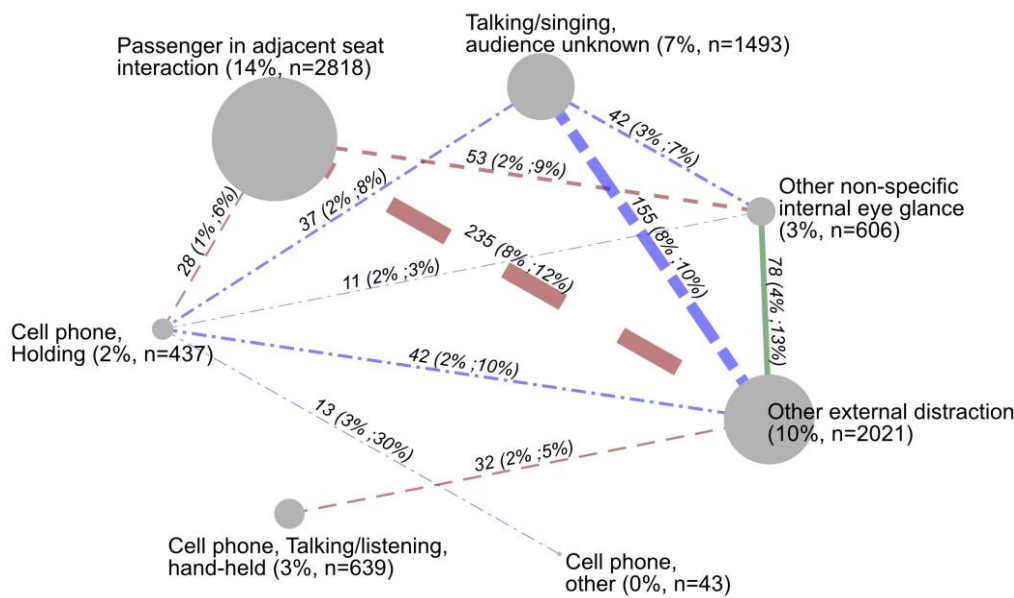

**Figure 4** Simplified co-occurrence graph for control driving segments with the following cutoff rules: 5 most prevalent events; co-occurrences with frequency ≥ 10 or co-occurrences with relative prevalence ≥ 30%.

The corresponding figure for Level 1-3 crashes is presented in Figure 5 below. In this case, due to the smaller number of events, co-occurrences of frequency ≥ 3 are shown.

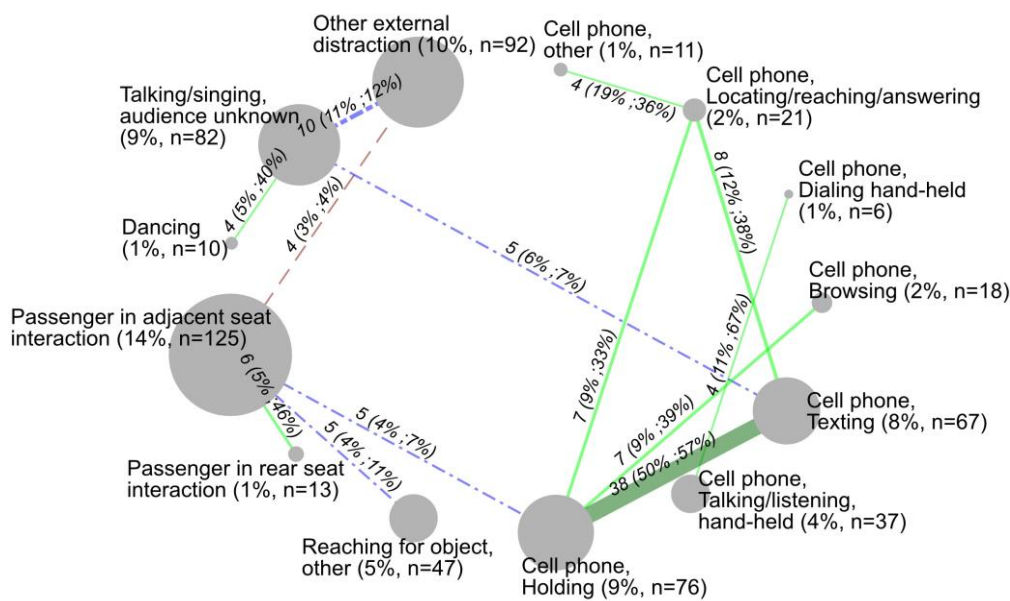

**Figure 5** Co-occurrence graph for level 1-3 crashes (n=891) with the following cutoff rules: 6 most prevalent events; co-occurrences with frequency ≥ 3 or co-occurrences with relative prevalence ≥ 30%.

## 3.3 Visual comparison of event types

The graphs displayed in Figure 6 below were constructed for the comparison of the main event types as specified in Section 2.1.1. This figure shows four co-occurrence graphs with the same set of nodes for each event type considered (CDS, CNC, L1-3 and rear-end striking) and in the same positions. In Figure 6, the graphs are presented side-by-side for easier comparison, but in a small size; full-sized graphs and the specification of cutoff parameters used to generate these graphs are provided in the Appendix.

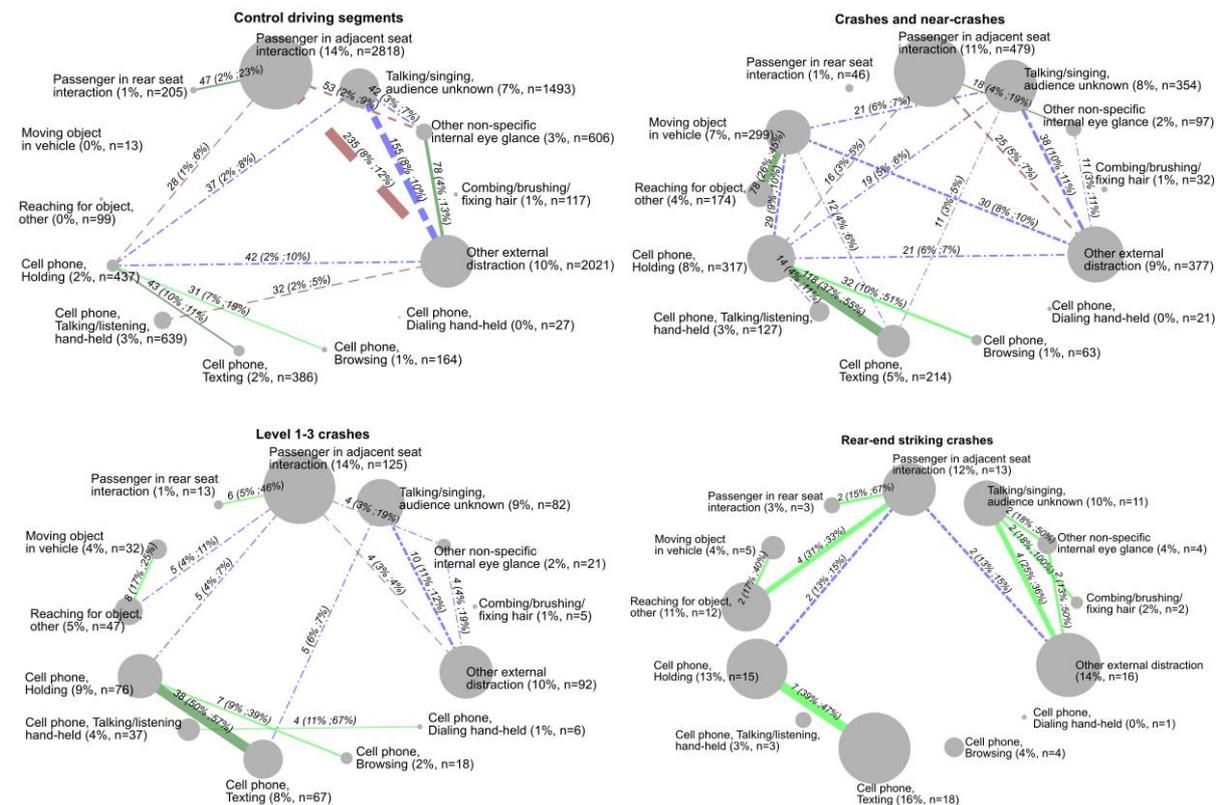

**Figure 6 Comparison of the secondary task co-occurrence structure of the four main event types considered. Green (solid) edges indicate greater frequency, red (dashed) edges indicate smaller frequency than independent co-occurrence, and blue (dash-dotted) edges indicate no significant difference compared to independent co-occurrence.**

Several observations can be made by a simple visual inspection of these graphs. For example, the nodes "Cell phone, holding" and "Cell phone, texting" (which are the end-nodes of a solid/green edge in the lower left corner of all graphs) are much larger in the graphs corresponding to safety-critical event types than in the graph representing control driving segments, and these tasks also co-occur much more frequently. Indeed, in more than 50% of Level 1-3 crashes when one of these tasks is present, the other one is present as well, in contrast to CDS where these secondary tasks are observed for 437 and 386 events respectively but they co-occur only 43 times, corresponding to 10% and 11% co-occurrence relative to the case counts (which is still significantly greater than what would be expected under a model of independent occurrence, as indicated by the dark green edge).

Similarly, "Reaching for object, other" and "Moving object in vehicle" (which are nodes on the left side of the graphs) are substantially more frequent among safety-critical event types than among control driving segments. For example, "Moving object in vehicle" which is essentially absent for CDS has 4% prevalence within level 1-3 crashes as well as rear-end striking crashes and 7% prevalence within CNC.

It is also interesting to compare the line styles/colors of edges in the graph: for example, the edge between "Passenger in adjacent seat – interaction" and "Other non-specific internal eye glance" is dark green (solid) in the graph for crashes and near-crashes and dark red (dashed) in the graph for control driving segments. This indicates that these secondary tasks tend to co-occur in CNC but tend not to co-occur in CDS. These tasks do not co-occur in any of the 112 rear-end striking crashes and their co-occurrence is not significantly different from what would be expected in a model based on independent occurrence within level 1-3 crashes, as indicated by the blue (dash-dotted) edge in the corresponding graph.

## 3.4 Risk estimates

The odds ratios for SAD and MAD compared to the "No Tasks" reference level, i.e., driving without any secondary tasks, computed for the different event types and using both sets of task definitions are shown in Figure 7 below. As in Section 3.1, the event types are given in ascending order from left to right by the prevalence of MAD. The highest odds ratios for both task definitions are $\widehat{OR}^{MAD/No\ Tasks}_{Rear-end\ striking}$ with corresponding confidence intervals of [5.11, 14.07] for SHRP2 task definitions and [4.04, 11.94] for G14 tasks; part of these confidence intervals is not visible in Figure 7 as the y-axis is shown from 0 to 9 only for a greater level of detail in this interval.

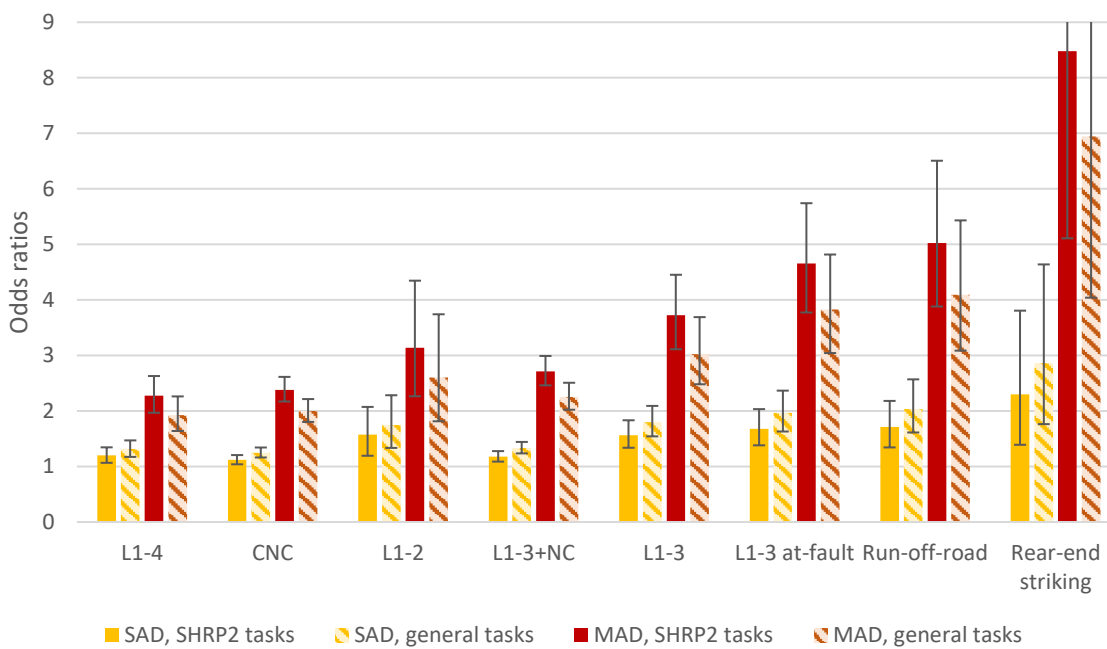

Figure 7 Comparison of odds ratios across event types for different task definitions, with "No Tasks" as reference level.

Table 4 below specifies the odds ratios and 95% confidence intervals for SAD, MAD and engagement in at least three tasks based on SHRP2 task definitions across event types.

Table 4 Odds ratios and confidence intervals comparing SAD, MAD and 3+ tasks to driving without secondary task engagement, SHRP2 task definitions.

|  | SAD | | MAD | | 3+ tasks | |
| --- | --- | --- | --- | --- | --- | --- |
|  | $\widehat{OR}$ | CI | $\widehat{OR}$ | CI | $\widehat{OR}$ | CI |
| L1-4 | 1.20 | [1.07, 1.35] | 2.27 | [1.97, 2.63] | 3.58 | [2.72, 4.71] |
| CNC | 1.12 | [1.04, 1.21] | 2.38 | [2.17, 2.61] | 4.42 | [3.70, 5.28] |
| L1-2 | 1.57 | [1.19, 2.07] | 3.14 | [2.26, 4.35] | 6.05 | [3.56, 10.29] |
| L1-3+NC | 1.18 | [1.09, 1.28] | 2.71 | [2.46, 2.99] | 5.18 | [4.32, 6.21] |
| L1-3 | 1.57 | [1.34, 1.83] | 3.72 | [3.11, 4.45] | 6.31 | [4.64, 8.60] |
| L1-3 at-fault | 1.68 | [1.38, 2.03] | 4.65 | [3.77, 5.74] | 7.26 | [5.07, 10.38] |
| Run-off-road | 1.71 | [1.34, 2.18] | 5.03 | [3.88, 6.51] | 7.07 | [4.52, 11.07] |
| Rear-end striking | 2.30 | [1.39, 3.81] | 8.48 | [5.11, 14.07] | 16.53 | [8.15, 33.52] |

According to the results in Table 4, for any event type X, the odds ratios are ordered as follows:

$$\widehat{OR}_X^{SAD/No\ Tasks} < \widehat{OR}_X^{MAD/No\ Tasks} < \widehat{OR}_X^{3+\ tasks/No\ Tasks}$$

and none of the confidence intervals contains 1. The same conclusions hold for the odds ratios based on G14 tasks (for which the table containing the corresponding ORs and CIs is provided in the Appendix).

## 4 DISCUSSION

### 4.1 Approach

This study analyzed the prevalence of drivers' engagement in multiple secondary tasks additional-to-driving (MAD), explored the structure of MAD, and quantified the associated risk, based on naturalistic driving data. Therefore, this paper is one of the few studies that is explicitly focused on multiple driver distractions and is based on real-world driving data, providing new information regarding this safety issue on the roads. Specifically, data from the SHRP2 NDS was analyzed to quantify MAD during normal driving and safety-critical events of different severity and type. Point estimates and confidence intervals were also computed for the corresponding odds ratios estimating the risk associated with MAD compared to no task engagement. Additionally, a novel visual representation of multitasking was developed to indicate which secondary tasks co-occur frequently and which ones do not within the different event types.

One potential issue with analysis of MAD is that its definition depends on the specification of secondary tasks. For example, if two tasks co-occur in a substantial proportion of events where either is present, one could argue that they are part of a single, more general task. Solid (green) edges in our graphs for the visualization of multitasking may help identifying secondary tasks that naturally co-occur. However, if the line style (color) of an edge between two secondary tasks differs across event types, it may be argued that combining those tasks is not reasonable, because it may hide driver behavior changes across event type. The investigation of the most common secondary task co-occurrences in SHRP2 in Flannagan et al. (2018) shows that there is no strong evidence for tasks to be redefined as being the same for the purpose of multitasking analysis. Nevertheless, in order to investigate the sensitivity of results on task definitions, all analyses related to prevalence and odds ratios in this study were conducted with both the original task definitions and a set of very general secondary tasks (G14, see Table 1), defined for the specific purpose of a sensitivity analysis.

In the study, it was found that MAD occurs in 11% of control driving segments, 22% of CNC, 26% of level 1-3 crashes, and 39% of rear-end striking crashes with the SHRP2 task definitions and 9%, 16%, 17%, and 28%, respectively for the same event types if defined in terms of general task groups. The difference between prevalence results for different task definitions is smaller for control driving segments than for safety-critical event types. These results could potentially reflect the importance of subtasks within the same group for crashes, for example due to the possibly longer duration of the resulting distraction. Alternatively, the same phenomenon could also be observed if data coders were prone to code more subtasks within the same group for safety-critical events than for control driving segments (in spite of quality control procedures in place to alleviate this risk). However, irrespective of task definitions, MAD and especially engagement in three or more tasks was most prevalent in rear-end striking, run-off-road, and at-fault crashes; at the same time, the prevalence of SAD varied little across event types with levels around 40% using SHRP2 tasks and 48% using general task definitions.

The structure of MAD was studied using co-occurrence graphs developed specifically for this study. A strength of the graphical representation for the visualization of MAD is that it gives an immediate picture of the most prevalent secondary tasks and co-occurrences, both in absolute terms and compared to the expected frequency under an independence model. For identifying the least common co-occurrences, one should consider both the red edges and the missing links in the graph as it is plausible that several such co-occurrences are below the corresponding cutoff value. The most common co-occurrences of secondary tasks vary substantially among event types; for example, "Passenger in adjacent seat – interaction" and "Other non-specific internal eye glance" tend to co-occur in crashes and near-crashes but tend not to co-occur in control driving segments.

The comparison across graphs for different event types (Figure 6) shows a clear difference between control driving segments and safety-critical event types that could be context-related (Tivesten & Dozza, 2015) and may support the extraction of matching baselines for OR analysis (Victor, et al., 2014; Klauer, Dingus, Neale, Sudweeks, & Ramsey, 2006). The higher prevalence of cell phone manual tasks in safety-critical event types – and in rear-end striking crashes specifically – is in agreement with previous research showing the importance of keeping eyes on the road, especially to avoid rear-end crashes (Victor, et al., 2014). The frequent co-occurrence of holding and texting in these event types suggests that the necessity for task interruption and visual time sharing during texting may make this task particularly risky (Tivesten & Dozza, 2014; Morando, Victor, & Dozza, 2019). MAD is particularly prevalent for rear-end striking crashes that are largely related to visual distraction. It is plausible that the more secondary tasks co-occur the more likely is that at least one of them would require drivers to look away from the road, suggesting that the effects of MAD on crash risk may be similar to the effect of visual distraction.

The odds ratios of MAD using SHRP2 task definitions and the corresponding 95% confidence intervals are 2.38 (2.17-2.61) for CNC, 3.72 (3.11-4.45) for Level 1-3 crashes and 8.48 (5.11-14.07) for rear-end striking crashes. The corresponding ORs using general task groups to define MAD are slightly lower at 2.00 (1.80-2.21) for CNC, 3.03 (2.48-3.69) for Level 1-3 crashes and 6.94 (4.04-11.94) for rear-end striking crashes. One possible explanation for the odds ratios with MAD being smaller for general task definitions is that multitasking within the same general task can potentially indicate task involvement with a longer time duration, which could elevate the associated risk more than involvement in different task groups. This, however, is only a hypothesis as the duration of tasks is not coded in the database and requires further investigation. Additionally, it should be noted that the confidence intervals for the odds ratios with MAD using different task definitions overlap for every event type; hence, the difference could potentially be due to chance only.

### 4.2 Limitations and future research directions

The definition of MAD in this paper is based on which secondary tasks are coded in the event window defined in Section 2.1.2. It is worth noting that complex secondary tasks are coded as a single task in the SHRP2 database, although they may include multiple actions influencing different aspects of the driver sensorimotor control. For instance, texting is a single secondary task that may include reading and writing, two different actions requiring visual and manipulation efforts in different amounts. Further, engagement in MAD does not necessarily mean simultaneous engagement in the coded secondary tasks. This is fully in line with the definition of MAD in the literature review in Lansdown, Stephens, & Walker (2015) as "more than one concurrently attempted Single Additional-to-Driving (SAD) task […] Further, MAD tasks were required to take place during broadly the same time period." However, at the end of Section 3.1 in the cited publication, it is added that "[…] the presumed common assumption that multiple tasks must be considered to impose, more or less, simultaneous demands […] may be overly conservative, and it may be more systematically consistent to consider multiple task demands as those that may occur either in a parallel or in a serial fashion." Therefore, MAD should be considered at least as generally as it is done here and the estimates for the frequency of MAD are not overestimating prevalence but could rather be considered as lower bounds.

All results in this study are based on SHRP2 NDS data, collected in the United States and may not be representative to drivers outside the U.S. Furthermore, as noted in Flannagan, Bärgman, & Bálint (2019), young drivers (aged 16-24) are substantially overrepresented and old drivers (of age 65+) are slightly overrepresented in SHRP2 compared to the general U.S. driver population. Therefore, the findings do not necessarily generalize to driving populations other than the participants in SHRP2. It would be valuable to perform similar analyses based on data from other sources to check the universality of the results.

For the visual representation of the structure of MAD, the topology of the resulting graphs depends on the parameters chosen to simplify the graph, and it will require future research to find the most efficient way of using such a visualization in the analysis of traffic safety data and exploit its full potential. A possible step in this direction would be the development of a tool where the cutoff thresholds for the number of displayed tasks or relative frequency of co-occurrences can be changed (e.g., with a slider), to study the change as a function of the parameters. Another potential improvement would be the implementation of features from other graphical software packages, including those in qgraph (Epskamp, Cramer, Waldorp, Schmittmann, & Borsboom, 2012; Asikainen, Hailikari, & Mattsson, 2018).

Finally, all ORs in this paper are crude odds ratios without adjustments and with "No Tasks" as reference level. Besides the general ORs considered in this study addressing all SHRP2 drivers and all environments, it would be of interest to analyze the prevalence and associated risk of MAD for specific subgroups of drivers (e.g. by gender or age group) or in specific driving environments (e.g. urban or motorway). ORs may also be used to quantify crash risk associated with co-occurrences of specific secondary tasks instead of the general concept of MAD which entails the co-occurrence of any two secondary tasks. A visual comparison of line widths in the crash and CDS co-occurrence graphs indicates the magnitude of such co-occurrence ORs, and the numerical values can be computed using the edge labels and the number of considered crash and CDS events. Additionally, it could be analyzed whether any compensation strategies are applied before initiating MAD tasks; this could include the analysis of vehicle kinematics before the event window used for coding task involvement, as well as reviewing the relevant video recordings of drivers. More details regarding these aspects and identifying the most common co-occurrences of secondary tasks additional-to-driving for specific driver groups would aid the development of better countermeasures.

## 5 CONCLUSIONS

The current study addressed the prevalence, structure and risk associated with multitasking additional-to-driving based on naturalistic driving data from SHRP2. While the prevalence of single task additional-to-driving is almost constant at 40-44% across all event types including control driving segments, the extent of MAD differs substantially, ranging from 11% within CDS to 39% within rear-end striking crashes. Using more general tasks for the definition of MAD results in slightly lower prevalence, but a similar pattern across event types. A graph representation was constructed that allows mapping task prevalence and co-occurrence within an event type as well as a comparison between different event types. The odds ratios of MAD indicate an elevated risk of all safety-critical events, with the greatest increase of the risk of rear-end striking crashes. The results confirm that the reduction of driving performance from MAD observed in simulator studies is manifested in real-world crashes as well.

# Appendix

**Control driving segments**

Figure 8 Comparison graph included in Figure 6 – control driving segments with the following cutoff rules: 3 most prevalent events; co-occurrences with frequency ≥ 20 or co-occurrences with relative prevalence ≥ 50%.

**Crashes and near-crashes**

Figure 9 Comparison graph included in Figure 6 – crashes and near-crashes with the following cutoff rules: 3 most prevalent events; co-occurrences with frequency ≥ 10 or co-occurrences with relative prevalence ≥ 50%.

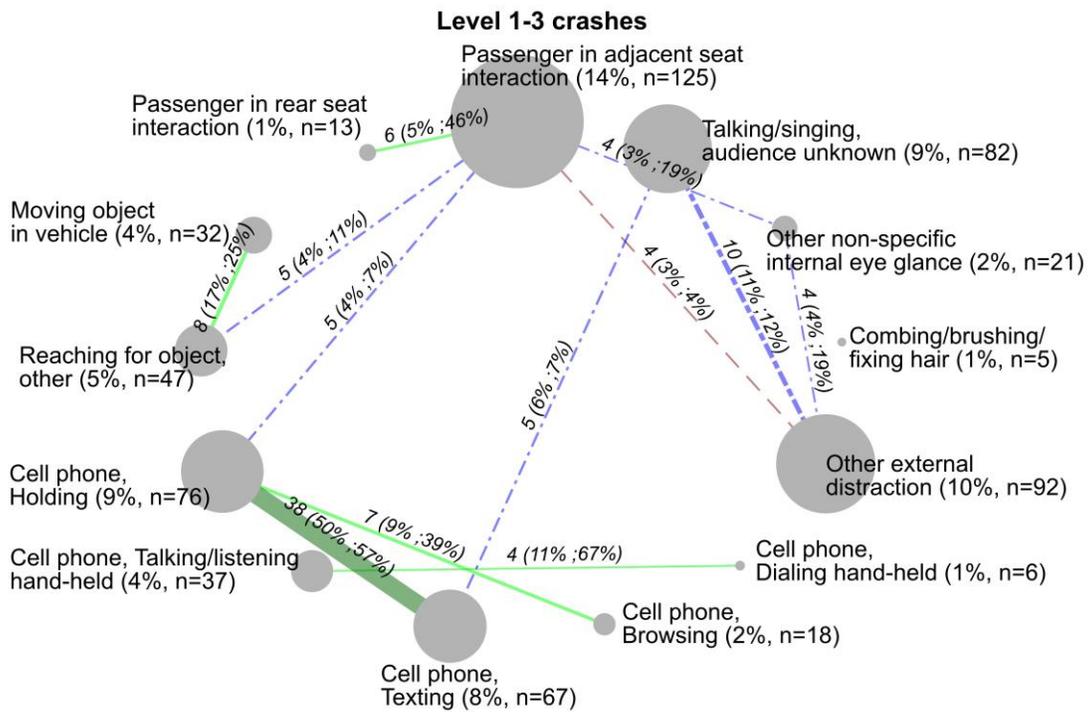

Figure 10 Comparison graph included in Figure 6 – level 1-3 crashes with the following cutoff rules: 3 most prevalent events; co-occurrences with frequency ≥ 3 or co-occurrences with relative prevalence ≥ 50%.

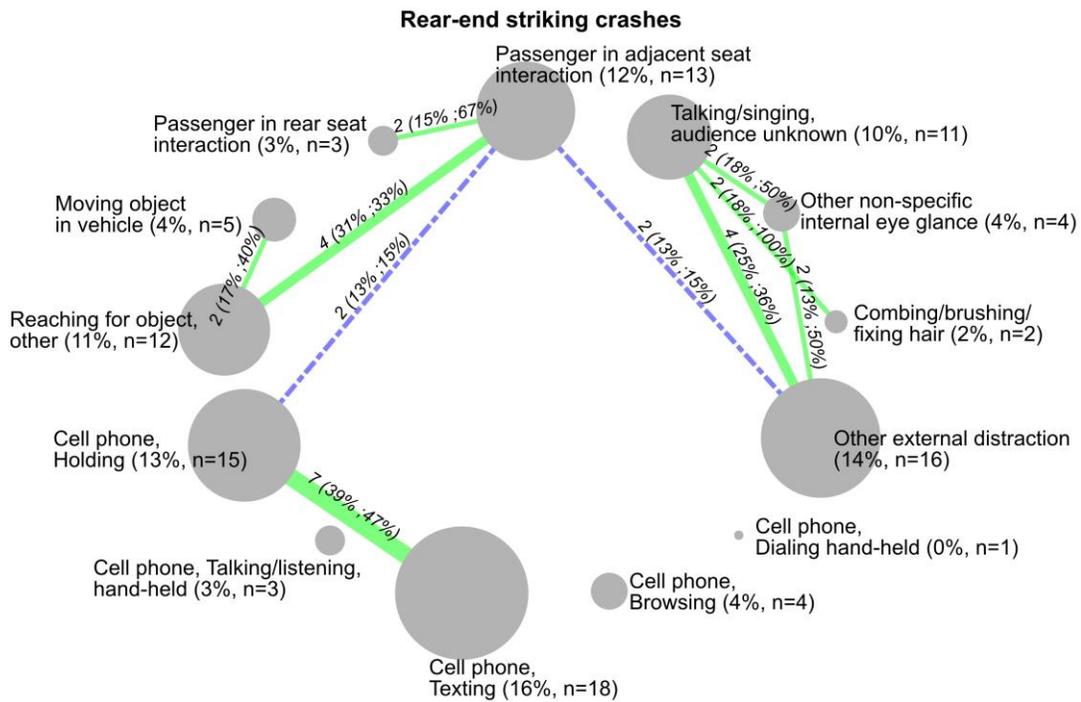

Figure 11 Comparison graph included in Figure 6 – rear-end striking crashes with the following cutoff rules: 3 most prevalent events; all co-occurrences.

Table 5 Odds ratios and confidence intervals comparing SAD, MAD and 3+ tasks to driving without secondary task engagement, G14 tasks.

|  | SAD | | MAD | | 3+ tasks | |
| --- | --- | --- | --- | --- | --- | --- |
|  | $\widehat{OR}$ | CI | $\widehat{OR}$ | CI | $\widehat{OR}$ | CI |
| L1-4 | 1.31 | [1.17, 1.47] | 1.93 | [1.64, 2.26] | 2.19 | [1.45, 3.31] |
| CNC | 1.25 | [1.16, 1.34] | 2.00 | [1.80, 2.21] | 2.13 | [1.62, 2.80] |
| L1-2 | 1.75 | [1.33, 2.28] | 2.61 | [1.81, 3.74] | 3.84 | [1.76, 8.41] |
| L1-3+NC | 1.34 | [1.24, 1.44] | 2.25 | [2.02, 2.51] | 2.30 | [1.73, 3.06] |
| L1-3 | 1.80 | [1.54, 2.09] | 3.03 | [2.48, 3.69] | 3.13 | [1.90, 5.15] |
| L1-3 at-fault | 1.96 | [1.63, 2.37] | 3.83 | [3.04, 4.82] | 3.82 | [2.18, 6.71] |
| Run-off-road | 2.04 | [1.61, 2.57] | 4.09 | [3.09, 5.43] | 3.49 | [1.68, 7.25] |
| Rear-end striking | 2.86 | [1.76, 4.64] | 6.94 | [4.04, 11.94] | 8.50 | [2.91, 24.81] |